\documentclass{article}

\usepackage{arxiv}

\usepackage[utf8]{inputenc} % allow utf-8 input
\usepackage[T1]{fontenc}    % use 8-bit T1 fonts
\usepackage{hyperref}       % hyperlinks
\usepackage{url}            % simple URL typesetting
\usepackage{booktabs}       % professional-quality tables
\usepackage{amsfonts}       % blackboard math symbols
\usepackage{nicefrac}       % compact symbols for 1/2, etc.
\usepackage{microtype}      % microtypography
\usepackage{lipsum}
\usepackage{fancyhdr}       % header
\usepackage{graphicx}       % graphics
\graphicspath{{media/}}     % organize your images and other figures under media/ folder
\usepackage{color}

\title{SuperPoint features in endoscopy
\thanks{\textit{\underline{Citation}}: 
\textbf{O.L. Barbed, F. Chadebecq, J. Morlana, J.M.M. Montiel, A.C. Murillo. SuperPoint Features in Endoscopy. MICCAI Workshop on Imaging Systems for GI Endoscopy, International Workshop on Graphs in Biomedical Image Analysis (2022). Pages 45-55. DOI:10.1007/978-3-031-21083-9\_5.}}
}

\author{
  O. L. Barbed, J. Morlana, J.M. Martínez Montiel, A. C. Murillo \\
  University of Zaragoza \\
  Zaragoza, Spain\\
  leon@unizar.es\\
   \And
  F. Chadebecq \\
  University College London \\
  London, UK\\
}

\begin{document}
\maketitle

\begin{abstract}
%This is a great paper and it has a concise abstract.
%This work explores the applicability of deep learning based image local feature extraction to endoscopy.  
There is often a significant gap between research results and applicability in routine medical practice. This work studies the performance of well-known local features on a medical dataset captured during routine colonoscopy procedures. 
Local feature extraction and matching is a key step for many computer vision applications, specially regarding 3D modelling. 
In the medical domain, handcrafted local features such as SIFT, with public pipelines such as COLMAP, are still a predominant tool for this kind of tasks. We explore the potential of the well known self-supervised approach SuperPoint~\cite{detone2018superpoint}, present an adapted variation for the endoscopic domain and propose a challenging evaluation framework. 
SuperPoint based models achieve significantly higher matching quality than commonly used local features in this domain. Our adapted model avoids features within specularity regions, a frequent and problematic artifact in endoscopic images, with consequent benefits for matching and reconstruction results. 
%Our results include a thorough analysis of different quality metrics in the proposed evaluation framework.
%We also include qualitative 3D reconstruction results on real colonoscopy data both with COLMAP and with our new proposed model. 
%Models and code available upon acceptance.
Training code and models available \url{https://github.com/LeonBP/SuperPointEndoscopy}
\end{abstract}

% keywords can be removed

\section{Introduction}
%%%WE WORK ON ENDOSCOPY
% Highlight the broad range of applications, put some "numbers" of patients and procedures performed per day?
Endoscopic procedures are a frequent medical practice. The endoscope guided by the physician traverses hollow organs or body cavities, such as the colon.
% Highlight the broad medical impact that improvements in these procedures can have
Improvements in quality and efficiency of this kind of procedures can benefit numerous patients and broaden screening campaigns reach. 
% Endomapper goals? 3D modeling for AR/VR 
In endoscopy, as in plenty other medical imaging tasks, computer vision has potential to help in numerous aspects, such as assistance for diagnosis~\cite{zhang2017mdnet} or 3D modelling~\cite{liu2020reconstructing}.
%{\color{red}
%Unfortunately, as in many other fields, 
Unfortunately, there is still a 
significant gap between research results and applicability into the clinic, as discussed for example in~\cite{chadebecq2020computer}. This study emphasizes the need for unsupervised methods that can fully exploit \textit{in the wild} medical data, which is in itself an already scarce resource. 
%}.  
To move forward, it is often key to consider challenging and realistic evaluations of current techniques, to determine where specific adaptations are needed.

%%%3D MODELS OF THE COLON
Our work is motivated by the automated acquisition of 3D models of the endoluminal scene, that can facilitate augmented reality applications or assistance for navigation or patient monitoring. A core step in 3D reconstruction techniques, such as structure from motion (SfM) or Simultaneous Localization and Mapping (SLAM), is local feature detection and matching. 
%%%WE FOCUS ON LOCAL FEATURE EXTRACTION
Many broadly used SfM or SLAM frameworks still rely on hand-crafted local feature computation~\cite{mahmoud2018live}, %widya2019whole,
% and medical domain tends to use this as well? 
although deep learning based techniques are boosting the state of the art. % in local feature detection and description. %and matching.
SuperPoint~\cite{detone2018superpoint} is one of the seminal works in this topic and has inspired many follow up works discussed next. %{\color{red}
This promising research stream of learning based local features is recently being exploited in the endoscopic image domain~\cite{liao2021deep},
%} %kist2021feature,
since evaluations and benchmarks on local feature detection and matching are typically focused on conventional images and mostly rigid scenes~\cite{jin2021image}.

\begin{figure}[!tb]
    \centering
    \footnotesize
    \begin{tabular}{@{}c@{\hspace{1mm}}c@{\hspace{1mm}}c@{\hspace{1mm}}c}
        \includegraphics[width=0.24\linewidth]{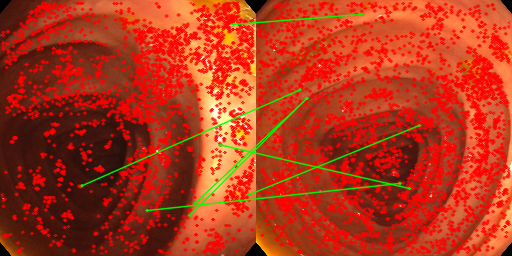} &
         \includegraphics[width=0.24\linewidth]{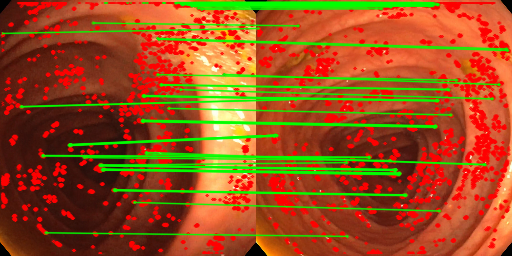} &
         \includegraphics[width=0.24\linewidth]{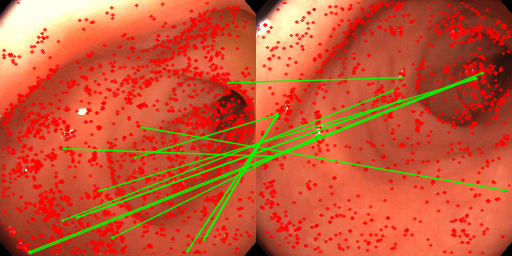} &
         \includegraphics[width=0.24\linewidth]{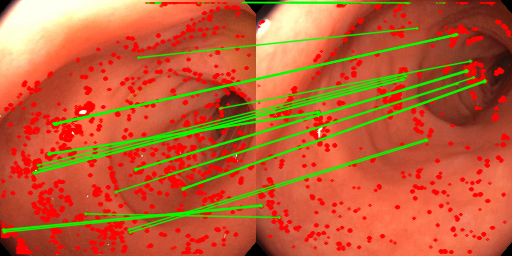}\\
         SIFT & Ours & SIFT & Ours\\
    \end{tabular}
    \caption{Feature extraction (red circle) and matching  (green line) on 
    endoscopy samples.
    %two pairs of 1 second apart frames from a real colonoscopy. General purpose hand-crafted features (SIFT) concentrate a lot on specularity artifacts, while our adapted SuperPoint achieves more and better distributed matches, key for good 3D reconstructions.
    }
    \label{fig:intro}
\end{figure}

%%%HARD DOMAIN
% Note the usually "challenging" quality of these recordings, specially if they are obtained from regular procedures, without any alteration of the doctors routines
Endoscopic images captured during routine procedures present many challenges (such as challenging textures, frequent artifacts and scene deformation) that hinder local feature extraction. Fig.~\ref{fig:intro} shows matches on two pairs of 1 second apart frames from a real colonoscopy where general purpose hand-crafted features (SIFT) can not tackle scenarios that our learned model features do. SIFT concentrates a lot on specularity artifacts, while our adapted SuperPoint model achieves more and better distributed matches, key for good 3D reconstructions. %matching results on two frames 1 second apart along a real endoscopy. General purpose hand-crafted features (SIFT in this case) concentrate a lot on specularity artifacts, while the adapted SuperPoint model in this work achieves more and better distributed correspondences. 
%%%CONTRIBUTION
The main contributions of this work are: 1) A thorough study of SuperPoint effectiveness on \textit{in the wild} endoscopic images, compared to typically used hand-crafted local features, including the proposed framework to evaluate these aspects in endoscopic data captured during daily medical practice; 2) our Superpoint adaptation to the endoscopic domain that improves its performance. %As previously mentioned, this work presents an evaluation framework to study the performance and possible adaptations of SuperPoint in endoscopy. 
%We selected SuperPoint as the base for our study since it is a seminal work, with numerous follow-ups, and typically a reference architecture when studying feature extractors.
%is motivated by the possibility to derive different training strategies allowing us to efficiently adapt this approach to endoscopy. Furthermore, SuperPoint is a fast method which could be efficiently integrated within surgical-navigation platforms and improve their robustness. 
%(c) make use of these local features in a downstream application of SfM in real endoscopy videos.

%We train the SuperPoint architecture with a heterogeneous set of real endoscopic sequences and propose a simple but effective modification to account for common endoscopic image artifacts due to illumination. We compare the quality of the features obtained with these models and existing methods on a novel real dataset captured during real colonoscopy procedures. 

%It is broadly accepted that to evaluate feature usefulness or applicability for final tasks, metrics on the actual final task are preferred to theoretical feature quality metrics such as repeatibility. Since there is no 3D reconstruction ground truth in the available real data, we can not measure the quality of the features for the final 3D reconstruction task, so we propose a set of metrics that account for amount, quality and dispersion of the features and the correspondences of the different local features studied. 

%Our quantitative and qualitative results demonstrate the effectiveness of the two new models trained, which are being released for the community, and the promising benefits that deep learning based features can bring to our target application in real colonoscopy data.

\section{Related Work}

% Two-Stream Deep Feature Modelling for Automated Video Endoscopy Data Analysis
% https://arxiv.org/abs/2007.05914
% IT's a MICCAI 2020.
% ~\cite{gammulle2020two}}
% larger field of view when building "local maps", 

% STILL USE hand crafted local features:
% ORB-slam hamlyn by Monti \cite{mahmoud2018live}
% https://ieeexplore.ieee.org/abstract/document/8410942?casa_token=RWgvrT2XOf8AAAAA:hPqg_NNVI0MaNoQIthNnPV7Gce2IB_thQIfSYwYUtbOsxU3FplAj4EZJNwypfT047CBQfDxjug
% SIFT 3d reconstruction stomach \cite{widya19whole}
% https://ieeexplore.ieee.org/abstract/document/8876662
%
% others that do not use local features but dense depth estimation IN ENDOSCOPY:
% paper Recasens
% paper Cipolla:  https://link.springer.com/chapter/10.1007%2F978-3-030-59716-0_74

% small paragraph intro "Machine learning in medical images"
%\subsection{3D modeling in intracorporeal images}
% applications (medical) that need feature matching: recognition and 3D reconstruction

% in particular ... endoscopic.
% comment hard problems (deformation (BARTOLI ETC), lack of texture (classic CV problem) ...) and potential uses, VR, help diagnosing ...

%[fr: It could be moved/integrated to the introduction to save space and if so we can add application exemples.]} 
\textbf{Endoscopic image registration for 3D reconstruction and mapping in mi\-ni\-ma\-lly-invasive surgery.}
%%%IMAGE REGISTRATION NOT WELL RESEARCHED IN ENDOSCOPY
Endoscopic image registration is an open problem essential to image-guided intervention. %It is particularly relevant to diagnostic and therapeutic endoscopic navigation which requires the localization and mapping of endoscopic scenes. Image registration
%This problem has been extensively studied, but it remains an active computer vision research problem and a challenging task in endoscopy, due to the paucity of reliable scene landmarks, lack of texture, complex tissue reflectance properties and the constrained manipulation of endoscopes within confined and deformable environments. Recent work 
%presents an evaluation benchmark to facilitate the development of novel techniques in this challenging domain~\cite{borgli2020hyperkvasir}.
%However, it remains an open problem with 
Current efforts are directed at developing benchmarks and techniques able to tackle this challenging domain~\cite{borgli2020hyperkvasir}. 
%%%VERY FEW LEARNING METHODS IN THIS DOMAIN
Learning-based approaches have shown their efficiency for general image registration, but they remain difficult to adapt to minimally-invasive imaging constraints, largely due to a lack of robust feature detection and matching in these scenarios. %Recent feature detection developments and benchmarking are mostly focused on domains far from endoscopy imagery. %such as endoscopy. 
Most vision-based approaches for 3D reconstruction in medical domains still rely on hand-crafted features \cite{gomez2021sd,espinel2021using}.  
%Such approaches have been used in different endoscopic datasets, but they suffer from a lack of robustness, predominantly because of their low-level local nature that does not embed contextual information.
%%%LEARNING METHODS HAVE GOOD QUALITIES FOR THIS PROBLEM
Some works 
%State-of-the-art learning-based endoscopic localization and mapping pipelines 
avoid the need for image registration by directly estimating and fusing key frames depth map \cite{ozyoruk2021endoslam} or combining them with camera pose estimates \cite{ma2019real}. 
%These methods demonstrate their efficiency on various endoscopy video sequences. 
These pipelines get the input frames in a temporally consistent way. Our work is focused on a more general problem of feature extraction without any temporal information given to the model. 

%We do not depend on supervised learning 
%they are only adapted to non-deformable scenes supervised learning 
%based on Structure-from-Motion \cite{hartley2004multiple},
%or implicitly assuming rigid photo consistency constraints)
%which limits the generalization.
%to other application numerous minimally-invasive procedures.

\textbf{Local feature detection and description for image registration.}
%%%OVERVIEW
%Most feature based endoscopy registration approaches rely on hand-crafted features, while learning-based registration methods have been mainly studied in the context of multi-modal image fusion. 
%This section is focused on 
%A survey on 
Image registration in general settings is a long studied problem \cite{ma2020image}. A key aspect in this work is 
learning-based methods for image registration in endoscopy. \\
%%%EARLY FEATURE DETECTION
\indent Early learning-based approaches solely focused on feature description. 
%Si\-a\-mese networks have notably been successfully combined with central-surround network for improving feature matching \cite{Zagoruyko15learning,tian17L2net}. 
Advanced training loss and strategies significantly improved feature descriptor performances, e.g., by relying on triplet loss which aims at maximizing descriptor discrepancy between close but negative pairs of matches %\cite{Balntas2016learning,Mishchuk2017working}.
\cite{Mishchuk2017working}.
Similar results have been achieved by sampling more negative pairs as proposed in \cite{tian17L2net}. %The latter approach further introduced the L2-Net architecture which has then been widely adopted for feature description. 
The learning-based feature detection problem has been less investigated. Former approaches learn to detect co-variant features and aim at reproducing and eventually improving hand-crafted feature detectors %extracting landmarks that are invariant to change of viewpoint 
\cite{di2018kcnn,laguna2019keynet}. Unlike these approaches, \cite{savinov2017quadnetwork} learns in an unsupervised way to rank keypoints according to their repeatability. The repeatability constraint is now generally combined with peakiness constraints for improving the robustness of the detector \cite{mishkin2018repeatability,zhang2018learning}. \\
%%%RECENT DETECTION+DESCRIPTION
\indent State-of-the-art registration approaches directly integrate feature extraction and description in a single framework. It has been shown that such approaches significantly improve matching results over classical hand-craf\-ted feature-based registration methods \cite{ono2018lf}. Preliminary approaches such as LIFT \cite{yi2016lift} aim at reproducing the different stages of classical image registration pipelines. %including the estimation of feature orientation. 
The need for Structure-from-Motion labels to train supervised methods makes these approaches impractical for applications such as endoscopy. More recent unsupervised approaches such as SuperPoint aim at jointly detecting and describing image landmarks \cite{detone2018superpoint}. %SuperPoint proposes a network composed of an encoder and two decoders generating a detector and a descriptor map. 
The training is an iterative process that starts by learning from a synthetic dataset of random 2D shapes. %where the landmark supervision is hand-crafted. 
The next iterations learn from the problem-specific dataset generated by applying random homographies to source images and using self-supervision from the previous iteration. It remains among the most efficient feature detection and description methods, and is still being considered in recent comparatives~\cite{jin2021image}.
%extracting landmarks using classical hand-crafted feature detectors. Despite the proposed method introduces a strong feature detection bias,  
The R2D2 network \cite{revaud2019r2d2}, based on the L2-Quad architecture, jointly estimates a reliability and repeatability map together with a dense descriptor map. Despite their efficiency, methods jointly detecting and describing features are difficult to train and do not generalize well to different application domains \cite{jin2021image}. To overcome these limitations, \cite{tian2020d2d} propose to rely on a describe-to-detect strategy which takes advantage of the efficiency and performances of learning-based descriptor models. %There exist few learning-based registration approaches adapted to endoscopy. 
Recently, \cite{liu2020extremely} propose a descriptor training strategy based on the formulation of a novel landmark tracking loss. While results demonstrate the efficacy of the proposed method, its computational cost remains generally high.
Recent image matching trends propose dense matching as an intermediate step to local matching~\cite{zhou2021patch2pix} and incorporating attention for the matching stages~\cite{sarlin2020superglue,jiang2021cotr,sun2021loftr}. However, these approaches rely on 3D reconstruction ground truth for training, which is  often not available for recordings acquired during routine medical practice.

\section{SuperPoint in endoscopy}
\label{sec:method}
%\section{Endoscopy SuperPoint}

Local feature matching is typically divided in four steps: feature detection, descriptor computation, matching and, often, outlier filtering. Our goal is to evaluate and improve existing methods on the first two steps for \textit{in the wild} endoscopy imagery. 
%As previously mentioned, 
The well-known SuperPoint, a seminal work regarding self-trained deep learning solutions for feature detection and description, is the base for our study. 
%The supplementary material includes implementation details on the methodology followed to apply SuperPoint in endoscopic images (Section~\ref{sec:implementation_details}). 
We next describe the Superpoint model variations used and the matching strategy applied.
More implementation details in the supplementary material.

% \paragraph{Image pre-processing.}
% Original raw endoscopic images are pre-processed with a few simple steps to fit the type of input that the SuperPoint model takes: 
% crop to select the largest (centered) squared patch within the image,
% and color-space conversion to grayscale. 

% \paragraph{Feature extraction and description.}
% Input images are fed to the SuperPoint model which outputs two sets of results: 1) location of the features extracted (a list of keypoints defined by $\{x_i,y_i,score_i\}$, where $x_i$ and $y_i$ are the pixel coordinates of the $i$th keypoint in the image, and $score_i$ is the confidence of the algorithm on that keypoint); 2) a list with the descriptors associated to each keypoint. Each descriptor is a L2-normalized list of floating point numbers whose dimensionality is set to $256$ as recommended in \cite{detone2018superpoint}.

% SuperPoint models are typically trained with relatively low resolution images, since larger resolutions make the training process very resource demanding and not feasible in commodity GPUs. We train on 256$\times$256  images, but use test images of 1080$\times$1080.

\subsection{SuperPoint models considered}
\paragraph{SuperPoint Base.}
Original SuperPoint model \cite{detone2018superpoint}. For this and the following model, we use the implementation by \cite{jau2020deep}%\footnote{\scriptsize \url{https://github.com/eric-yyjau/pytorch-superpoint}}
, which allows us to use the original model weights as well as training new models. 
SuperPoint follows the known encoder-decoder architecture, but with two parallel decoders (detection and description heads). SuperPoint processes a single image ($I \in \mathbb{R}^{H \times W}$. $H$ and $W$ are the height and width, respectively) as input and produces two outputs: \textit{detection}, image location of each keypoint extracted, and \textit{description}, one descriptor for each keypoint. 
The \textit{detection} head maps $I$ into a tensor $\mathcal{X} \in \mathbb{R}^{H/8 \times W/8 \times 65}$. The depth of $65$ corresponds to a cell of 8$\times$8 pixels in $I$ plus an additional channel called dustbin or ``no interest point''. After performing a softmax over the third dimension (we refer to it as $\mbox{softmd}()$), the dustbin is removed and the rest is reshaped to recover $I$'s dimensions ($\mbox{d2s}(X): \mathbb{R}^{H/8 \times W/8 \times 64} \to \mathbb{R}^{H \times W}$). The result is interpreted as a probability heatmap of the keypoints in the image. 
The \textit{description} head maps $I$ into a tensor $\mathcal{D} \in \mathbb{R}^{H/8 \times W/8 \times 256}$. The depth of $256$ is the descriptor size, associated with a whole cell of 8$\times$8 pixels in $I$. Bi-cubic interpolation is used to upsample $\mathcal{D}$ into having $H$ and $W$ as the first two dimensions. The descriptors are L2-normalized. 
SuperPoint is trained by contrasting the outputs of an image and a warped version of itself via a known homography and pre-computed pseudo-labels of image keypoints. The \textbf{loss function} is
\begin{equation}
\small
    \mathcal{L}_{SP} \left( \mathcal{X}, \mathcal{X}^{\prime}, \mathcal{D}, \mathcal{D}^{\prime} ; Y, Y^{\prime}, S \right) = 
    \mathcal{L}_{p} \left( \mathcal{X}, Y \right)
    +\mathcal{L}_{p} \left( \mathcal{X}^{\prime}, Y^{\prime} \right)  
    +\lambda \mathcal{L}_{d} \left( \mathcal{D}, \mathcal{D}^{\prime}, S \right),
    \label{eq:LSP}
\end{equation}
\noindent where $\mathcal{X}$ and $\mathcal{X}^{\prime}$ are the raw detection head outputs for image $I$ and warped image $I^{\prime}$, respectively. Their associated detection pseudo-labels are $Y$ and $Y^{\prime}$. $\mathcal{D}$ and $\mathcal{D}^{\prime}$ are the raw description head outputs. $S \in \mathbb{R}^{H/8 \times W/8 \times H/8 \times H/8}$ is the homography-induced correspondence matrix. $\mathcal{L}_p$ is the detection loss, which measures the discrepancies between the detection outputs and the pseudo-labels. $\mathcal{L}_d$ is the description loss, that forces descriptors that correspond to the same region in the original image to be similar, and different to the rest. $\lambda$ is a weighting parameter.

\paragraph{E-SuperPoint.}%\paragraph{E-SuperPoint+S.} 
Specularities are very frequent artifacts in endoscopic images~\cite{stoyanov2005removing}, and feature extractors often tend to detect features in the contour or within these image specularities. %which are very harmful for many image processing tasks
Features on specularities are not well suited for rigid model estimation, suffer from bad localization, and they turn out to be unreliable in downstream tasks such as tracking and 3D reconstruction. Although they can be masked out later, as we see in our experiments, they  account for a too large portion of the features and matches. Thus, we aim to prevent them from happening in the first place, to encourage the detectors to focus on other regions. 

We fine-tune the original model using endoscopic images (resized to 256$\times$256) from routine medical practice recordings (dataset detailed in Sec.~\ref{sec:experiment}).
%The training data, detailed in next section~\ref{sec:data}, consists of $125000$ endoscopy images captured during real practice in two different hospitals. Besides, we save $7179$ frames for validation. 
Pseudo-labels are obtained % via Homographic Adaptation ($N_h=100$) 
with the original SuperPoint model. %We average the detection over $100$ homography transformations of each image.
The pseudo-label is set to zero where the confidence value is lower than a threshold of $0.015$. Non-maximum suppression is applied over windows of 9$\times$9 pixels, and only the top $600$ points are finally saved. 
We fine-tune the model for $200000$ iterations with learning rate of $1\mathrm{e}{-5}$ and batch size of $2$. We use  sparse loss for more efficient convergence~\cite{jau2020deep}, and the rest of  parameters are the same as they describe. 
For testing we set the detection threshold to $0.015$ and non-maximum suppression over 3$\times$3 windows. 
Our modification of the SuperPoint model adds a new term to the training loss, our specularity loss $\mathcal{L}_s$. %, applied during the fine-tuning of the model. % to accomplish this goal
The purpose of $\mathcal{L}_{s}$ is to account for all the keypoints that are extracted on top of specularities, and is close to zero when there are no keypoints on those locations.
% This is the \textbf{original SuperPoint loss} function:%
% \begin{equation}
% \small
%     \mathcal{L}_{SP} \left( \mathcal{X}, \mathcal{X}^{\prime}, \mathcal{D}, \mathcal{D}^{\prime} ; Y, Y^{\prime}, S \right) = 
%     \mathcal{L}_{p} \left( \mathcal{X}, Y \right)
%     +\mathcal{L}_{p} \left( \mathcal{X}^{\prime}, Y^{\prime} \right)  
%     +\lambda \mathcal{L}_{d} \left( \mathcal{D}, \mathcal{D}^{\prime}, S \right),
%     \label{eq:LSP}
% \end{equation}
% \noindent where $\mathcal{L}_p$ is the detection loss, $\mathcal{L}_d$ the description loss, and $\lambda$ is a weighting parameter. $\mathcal{X}, \mathcal{X}^{\prime} \in \mathbb{R}^{H/8 \times W/8 \times 65}$ are the raw detection head outputs for image $I \in \mathbb{R}^{H \times W}$ and its warped pair $I^{\prime} \in \mathbb{R}^{H \times W}$, respectively. Their associated detection pseudo-labels are $Y$ and $Y^{\prime}$. $\mathcal{D}, \mathcal{D}^{\prime} \in \mathbb{R}^{H/8 \times W/8 \times 256}$ are the raw description head outputs. $S \in \mathbb{R}^{H/8 \times W/8 \times H/8 \times H/8}$ is the homography-induced correspondence matrix. 
\textbf{The final loss} is: 
% \begin{align}
%     \mathcal{L}_{ESP} (I, I^{\prime}, \mathcal{X}, \mathcal{X}^{\prime}, \mathcal{D}, \mathcal{D}^{\prime} &; Y, Y^{\prime}, S ) = \nonumber\\
%     \mathcal{L}_{SP}& ( \mathcal{X}, \mathcal{X}^{\prime}, \mathcal{D}, \mathcal{D}^{\prime} ; Y, Y^{\prime}, S )\nonumber\\
%     + \lambda_s \mathcal{L}_{s} &( \mathcal{X}, I )
%     + \lambda_s \mathcal{L}_{s} ( \mathcal{X}^{\prime}, I^{\prime} ),
% \end{align}
\begin{equation}
\small
    \mathcal{L}_{ESP} (I, I^{\prime}, \mathcal{X}, \mathcal{X}^{\prime}, \mathcal{D}, \mathcal{D}^{\prime} ; Y, Y^{\prime}, S ) = 
    \mathcal{L}_{SP} ( \dots )%\mathcal{X}, \mathcal{X}^{\prime}, \mathcal{D}, \mathcal{D}^{\prime} ; Y, Y^{\prime}, S )
    + \lambda_s \mathcal{L}_{s} ( \mathcal{X}, I )
    + \lambda_s \mathcal{L}_{s} ( \mathcal{X}^{\prime}, I^{\prime} ),
\end{equation}
\noindent where we add to the original $\mathcal{L}_{SP}$ the value of our specularity loss $\mathcal{L}_{s}$, once per image,  weighted by the scale factor $\lambda_s$. $\mathcal{L}_s$ is defined as
\begin{equation}
    \mathcal{L}_{s} \left( \mathcal{X}, I \right) = \frac{\sum_{h,w=1}^{H,W}\left[\mbox{m}\left(I\right)_{hw} \cdot \mbox{d2s}\left(\mbox{softmd} \left(\mathcal{X}\right)\right)_{hw}\right]}{\epsilon + \sum_{h,w=1}^{H,W}\mbox{m}\left(I\right)_{hw}},
\end{equation}
% \begin{align}
%     \mathcal{L}_{t} \left( \mathcal{D}_a, \mathcal{D}_b, M \right) = \frac{1}{P^2} \sum_{j=1}^{P} \sum_{j'=1}^{P} l_t \left( \mathbf{d}_{m_{aj}}, \mathbf{d}_{m_{bj'}}', j, j' \right) 
% \end{align}
% \noindent where
% \begin{align}
%     l_t \left( \mathbf{d}, \mathbf{d}', j, j' \right)& = \lambda_t * (j=j') * max(0,m_p - \mathbf{d}^T \mathbf{d}')\\
%     & + (j\neq j') * max(0,\mathbf{d}^T \mathbf{d}'-m_n).
% \end{align}
\noindent where $\mbox{softmd}()$ and $\mbox{d2s}()$ are softmax and reshape functions from the original SuperPoint,  % is the softmax function applied on the third dimension, %\\
%$\mbox{d2s}(X): \mathbb{R}^{H/8 \times W/8 \times 65} \to \mathbb{R}^{H \times W}$ is a flattening function that, as described in the original SuperPoint, eliminates the dustbin, 
and $\epsilon = 10^{-10}$. The subscript $X_{hw}$ refers to the value of $X$ at row $h$ and column $w$. 
% $\mbox{m}(I)_{hw}$ is $1$ if $I_{hw}>0.7$ and $0$ otherwise,  
% %\begin{align}
% %    m\left(I\right)_{hw} = \begin{cases}
% %        1, & \text{if } I_{hw}>0.7\\
% %        0, & \text{otherwise} 
% %        \end{cases},
% %\end{align}
% %\noindent which
% so it outputs a binary mask for the pixels in the image with high intensity value. 
$\mbox{m}(I)$ is a weighting mask: it is $>0$ for pixels near a specularity and $0$ otherwise. The mask comes from post-processing $I$ with three operations: a binary threshold of $I_{hw}>0.7$, a dilation of this binary output with a 3$\times$3 kernel size, and a Gaussian blur of the mask with 9$\times$9 kernel size and $\sigma=4$. 
The threshold of $0.7$ was chosen empirically, after observing that higher values missed too many specularities and lower values were discarding too many valid regions. 
%is the approximate value of our threshold $180$ after normalizing by the maximum intensity value of $255$. 
%The training parameters are the same as previous model. 
To balance the new loss component $\mathcal{L}_s$, we set the weighting parameter $\lambda_s=100$ so the losses have similar magnitudes for better optimization. Testing parameters remain the same.

\subsection{SuperPoint Matching}
SuperGlue \cite{sarlin2020superglue} is a well-known matching strategy proposed for SuperPoint. However, it requires correspondence ground-truth for training so we can not easily adapt it to endoscopy imagery. %Moreover, we want to evaluate different feature extractors in a fair setting and SuperGlue also requires retraining for other extractors. 
We opt to use bi-directional brute force matching, the originally recommended matching for SuperPoint. 
%all the other explored extractors. 
We also perform a robust geometry estimation with RANSAC to remove outliers, assuming local rigidity for short periods of time. Matching of frames too far apart along the video would need to account for significant deformations, which is out of the scope for this work.
%assuming that the endoscopic scene is almost rigid for small intervals (strong deformations typically occur in much larger time frames than the ones considered in this matching).  
%matches so the final set of inlier matches is much more accurate.
% In both cases, matches' coordinates are rectified using the camera intrinsics. 

\section{Experiments}
\label{sec:experiment}
This section summarizes the main results and insights from our comparison of different SuperPoint models and well-known local features applied in endoscopic data. % described in previous section.
Implementation details are in the supplementary material, Table 1.

\paragraph{Datasets.}
A key aspect in this research is to evaluate local feature performance on \textit{in the wild} endoscopic recordings. The model is trained on a set of private videos, and evaluated on two public benchmarks: EndoMapper~\cite{azagra2022endomapper} and Hyper-Kvasir~\cite{borgli2020hyperkvasir}. The supplementary material includes sample images from all sets. %, Figures 1 and 2.

%\begin{itemize}
    %\item 
    $\bullet$ \textbf{\textit{Train set}.} Endoscopy videos captured across several days of regular medical practice, each video corresponding to a routine procedure on a different patient. We use $11$ videos and extract $125000$ training frames and another $7179$ for validation.
    %From the first hospital we use $7$ videos ($57946$ training frames and $3316$ for validation). From the second one, $4$ videos ($67054$ train and $3863$ validation frames). 
    
    %\item 
    $\bullet$ \textbf{EndoMapper \textit{test set}}. $6$ full endoscopies for testing ($14191$ frames). Sequences 1, 2, 14, 16, 17 and 95. This dataset is similar to the videos used for training. 
    
    %\item 
    $\bullet$ \textbf{Hyper-Kvasir \textit{test set}.} $31$ short test videos (total of $51925$ frames). The labeled videos in ``lower-gi-tract/quality-of-mucosal-view/BBPS-2-3''.
\paragraph{Evaluation framework proposed.}
%A motivation to obtain better local features in endoscopic imagery is to boost 3D modeling strategies in this domain. 
As  often discussed in recent literature, common matching quality metrics, such as repeatibility or homography estimation, are not fully representative of local features behaviour in real world settings~\cite{jin2021image}. 
%However, metrics suggested as better alternatives typically rely on ground truth data to allow the computation of motion estimation  or 3D reconstruction errors. 
%A common evaluation challenge with real data is the lack of ground truth of any kind. For our endoscopic video data from real procedures, we have no correspondences, nor motion estimation ground truth.
%For reference, if we evaluate local features quality with the well known Homography Estimation Score, as suggested in \cite{mikolajczyk2005performance} and also used for instance in \cite{detone2018superpoint}, in our two evaluation subsets SIFT obtains the highest accuracy, while SuperPoint obtains significantly lower scores. This score represents the accuracy of estimated homographies computed over a set of pairs. The pairs are generated artificially to have a ground truth homography to compare with.
This work also shows that hand-crafted features, particularly SIFT, can still surpass more recent deep learning based features regarding accuracy in 3D vision tasks such as image registration. 
%They also note that the evaluation based only on theoretical feature quality like repeatability metrics can be misleading.}
For features and matches to be useful in posterior 3D reconstruction tasks, known desired properties include: good amount of quality matches (reliable and accurate) and matches covering all the scene to better capture the 3D scene information. We propose the following for the evaluation:

$\bullet$ To use an \textbf{existing SfM approach}, COLMAP~\cite{schoenberger2016mvs,schoenberger2016sfm}, to \textbf{pre-compute a pseudo-ground truth} for the relative pose between each pair of frames. %{\color{red}In the supplementary material, Section~\ref{sec:3d_colmap} shows an example of the reconstruction capabilities of this software in endoscopy scenes.}
%This state-of-the-art SfM pipeline includes a
COLMAP runs a final global bundle adjustment optimization to recover all relative camera poses. This pseudo-ground truth is used to compute rotation estimation errors and matching quality metrics detailed next. %More details on how this process and pseudo-ground truth are included in the supplementary material, Section 1.

$\bullet$ A set of \textbf{matching quality metrics}  %(detailed in the supplementary material, Section 1, Table 1) 
to account for: 1) matching quantity and quality, with inliers obtained from Essential (when camera calibration is available, E Inl.) or Fundamental (F Inl.) matrix RANSAC-based estimation, and inliers according to the relative pose provided as pseudo-ground truth (pGT Inl.); 2) scene coverage, with image cell \% (out of a $16\times16$ grid) with at least one inlier  (\textit{\%Gr}). %As we can see next, 
\paragraph{Matching quality evaluation.}
The following experiments analyze how well each feature can be matched along challenging endoscopic sequences.
% MUEVO ESTO MAS ARRIBA, pero resumido
%First of all, there have to be enough matches so the process can extract some geometric understanding of the pair of images: the more matches to choose from, the more probability there are enough matches to estimate the geometry correctly. This implies that the quality of the matches is also important: the higher the number of matches that suggest the same geometry, the more probability it is correct and also more accurately estimated. However, quantity and quality by themselves suggest a convergence to the trivial solution of having a very large number of equivalent matches concentrated in a very small region of the image. To avoid this convergence, we also evaluate the methods in the dispersion of their matches. Since dispersion is hard to evaluate perfectly with a single metric, we chose two very distinct from one another.
%In this experiment, 
We extract and match features across pairs of frames $1$ second apart from each other from sequences in EndoMapper ($1$ second $= 40$ frames for three videos, $1$ s $= 50$ frames for the other three) and Hyper-Kvasir ($1$ second $= 25$ frames). %in two different scenarios: %1 and Eval2
\begin{table*}[!tb]
%\small
\footnotesize
\centering
    \begin{tabular}{@{}cc}
        \begin{tabular}{@{}|l|c|c@{\hspace{2mm}}c|c@{\hspace{2mm}}c|}
         \cline{2-6}
         %\multicolumn{1}{c|}{} & \textbf{Feat}  & \multicolumn{4}{c||}{\textbf{ s1 (Bi-NN Matching)}} & \multicolumn{4}{c|}{\textbf{ s40 (Bi-NN Matching)}}\\%  &
         \multicolumn{1}{c|}{} &  \textbf{Feat/Img} &  \textbf{E Inl.} & \%\textbf{\textit{Gr}} &  \textbf{pGT Inl.} & \%\textbf{\textit{Gr}} \\
         \hline
         % $2350.2$ & $0.0/230.1$ ($0.0\%$) & $nan$ & $nan$ & $148.2/230.1$ ($64.4\%$) & $0.268$ & $11.9$ & $124.3/230.1$ ($54.0\%$) & $0.268$ & $7.7$ & $71.6/230.1$ ($31.1\%$) & $0.253$ & $9.3$
         \textbf{SIFT} & $2350.2$ &  $148.2$ & $11.9$ &  $71.6$ & $9.3$\\
         % $2163.0$ & $0.0/1005.2$ ($0.0\%$) & $nan$ & $nan$ & $153.0/1005.2$ ($15.2\%$) & $0.260$ & $8.5$ & $118.6/1005.2$ ($11.8\%$) & $0.260$ & $5.8$ & $80.9/1005.2$ ($8.0\%$) & $0.247$ & $6.0$
         \textbf{ORB} & $2163.0$ &  $153.0$ & $8.5$ &  $80.9$ & $6.0$\\
         \hline
         % $1333.7$ & $0.0/577.8$ ($0.0\%$) & $nan$ & $nan$ & $96.4/577.8$ ($16.7\%$) & $0.268$ & $11.4$ & $79.1/577.8$ ($13.7\%$) & $0.265$ & $8.3$ & $57.1/577.8$ ($9.9\%$) & $0.259$ & $8.2$
         \textbf{SP Base} & $1333.7$ &  $96.4$ & $11.4$ &  $57.1$ & $8.2$\\
         % $4500.9$ & $0.0/1349.7$ ($0.0\%$) & $nan$ & $nan$ & $278.9/1349.7$ ($20.7\%$) & $0.295$ & $13.2$ & $197.5/1349.7$ ($14.6\%$) & $0.272$ & $11.0$ & $172.0/1349.7$ ($12.7\%$) & $0.294$ & $9.8$
         \textbf{E-SP} & $\textbf{4500.9}$ &  $\textbf{278.9}$ & $\textbf{13.2}$ &  $\textbf{172.0}$ & $\textbf{9.8}$\\
         \hline
         \multicolumn{6}{c}{(a) \textbf{EndoMapper} \textit{test set} (1080x1080 resolution)} \\
        \end{tabular} &
        \begin{tabular}{|c|cc|}
         \cline{1-3}
         \textbf{Feat/Img} &  \textbf{F Inl.} & \%\textbf{\textit{Gr}}\\% &  \textbf{s1: H Inl.} & \%\textbf{\textit{Gr}}
         \hline
         $\textbf{825.7}$ & $151.3$ & $\textbf{18.6}$\\ 
         $361.3$ & $137.2$ & $6.3$\\ 
         \hline
         $211.8$ & $51.3$ & $11.1$\\ 
         $591.3$ & $\textbf{200.4}$ & $11.3$\\
         \hline
         \multicolumn{3}{c}{(b) \textbf{Hyper-Kvasir} (512x512)}
        \end{tabular}
    \end{tabular}
    \caption{Matching quality metrics for the two different test sets. E-SP trained on \textit{Train set}. pGT only available for (a) EndoMapper data.
    %. Hospital-1 \textit{test set} (1080x1080  resolution)
    }
    \label{tab:inliers_all}
\end{table*}

Table~\ref{tab:inliers_all} %and~\ref{tab:inliers_hyperkvasir} 
shows the performance of different baselines and our adapted model. The changes proposed have a noticeable effect, obtaining improvements in amount of features extracted and inlier matches (both with RANSAC and with the pseudo-GT) and spreading of these matches over the image in both scenarios. This is remarkable because E-SP was not fine-tuned in (b) Hyper-Kvasir data but it still mostly outperforms the rest.

\begin{figure*}[!tb]
    \centering
    \footnotesize
    %\vspace{-0.5cm}
    \begin{tabular}{@{}c@{\hspace{1mm}}c@{\hspace{1mm}}c@{\hspace{1mm}}c}
    % \includegraphics[width=0.13\textwidth]{images/examples_features/sift_v33_s19_1_2.png}
    % & \includegraphics[width=0.13\textwidth]{images/examples_features/orb_v33_s19_1_2.png}
    % & \includegraphics[width=0.13\textwidth]{images/examples_features/sp_v33_s19_1_2.png}
    % & \includegraphics[width=0.13\textwidth]{images/examples_features/spft_v33_s19_1_2.png}
    % & \includegraphics[width=0.13\textwidth]{images/examples_features/spspec_v33_s19_1_2.png}\\
    % \includegraphics[width=0.13\textwidth]{images/examples_features/sift_v33_s19_1_40.png}
    % & \includegraphics[width=0.13\textwidth]{images/examples_features/orb_v33_s19_1_40.png}
    % & \includegraphics[width=0.13\textwidth]{images/examples_features/sp_v33_s19_1_40.png}
    % & \includegraphics[width=0.13\textwidth]{images/examples_features/spft_v33_s19_1_40.png}
    % & \includegraphics[width=0.13\textwidth]{images/examples_features/spspec_v33_s19_1_40.png}\\
    \includegraphics[width=0.24\textwidth]{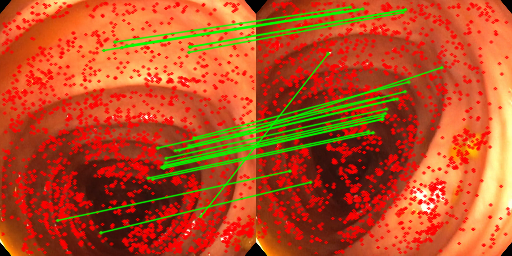}
    & \includegraphics[width=0.24\textwidth]{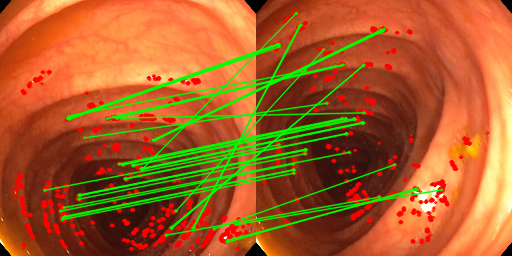}
    & \includegraphics[width=0.24\textwidth]{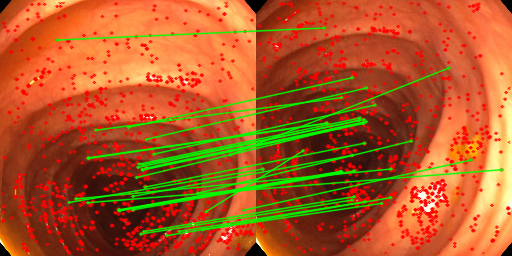}
    & \includegraphics[width=0.24\textwidth]{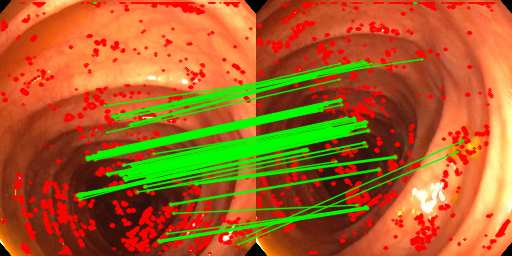}\\
    % & \includegraphics[width=0.15\textwidth]{images/examples_features/spspec_v33_s19_36_37.png}\\
    \includegraphics[width=0.24\textwidth]{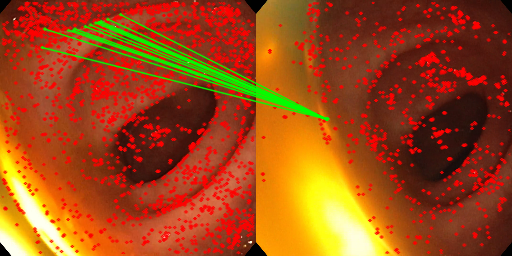}
    & \includegraphics[width=0.24\textwidth]{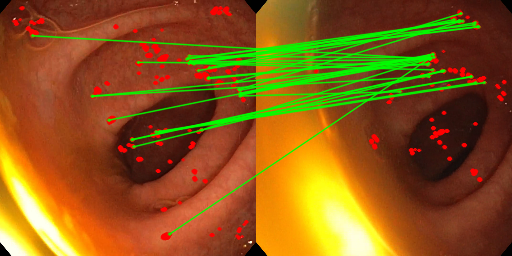}
    & \includegraphics[width=0.24\textwidth]{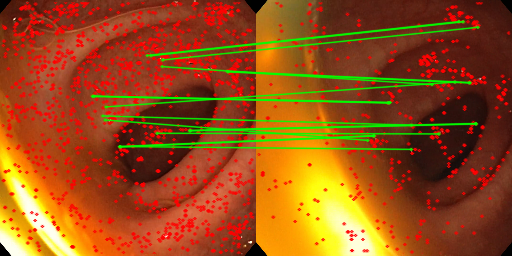}
    & \includegraphics[width=0.24\textwidth]{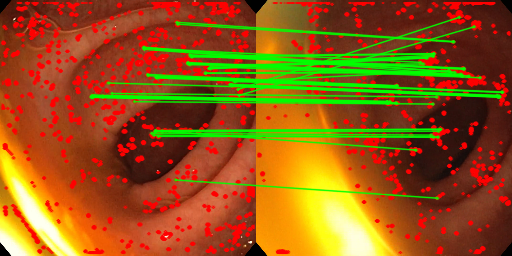}\\
    % & \includegraphics[width=0.15\textwidth]{images/examples_features/spspec_v33_s19_36_75.png}\\
    %\includegraphics[width=0.19\textwidth]{images/examples_features/sift_v34_s3_15_16.png}
    %& \includegraphics[width=0.19\textwidth]{images/examples_features/orb_v34_s3_15_16.png}
    %& \includegraphics[width=0.19\textwidth]{images/examples_features/sp_v34_s3_15_16.png}
    %& \includegraphics[width=0.19\textwidth]{images/examples_features/spft_v34_s3_15_16.png}
    %& \includegraphics[width=0.19\textwidth]{images/examples_features/spspec_v34_s3_15_16.png}\\
    %\includegraphics[width=0.19\textwidth]{images/examples_features/sift_v34_s3_15_54.png}
    %& \includegraphics[width=0.19\textwidth]{images/examples_features/orb_v34_s3_15_54.png}
    %& \includegraphics[width=0.19\textwidth]{images/examples_features/sp_v34_s3_15_54.png}
    %& \includegraphics[width=0.19\textwidth]{images/examples_features/spft_v34_s3_15_54.png}
    %& \includegraphics[width=0.19\textwidth]{images/examples_features/spspec_v34_s3_15_54.png}\\
    SIFT  & ORB & SuperPoint Base & E-SuperPoint
    \end{tabular}
    \caption{Features (red circles) and inlier matches after RANSAC (green lines) obtained for two pairs of 1 second apart frames using different features.}
    \label{fig:inliers1}
\end{figure*}

\paragraph{Specularities.}
E-SuperPoint is designed to encourage feature extraction avoiding specularities. This experiment evaluates this with the number of features and inliers when features located in specularity pixels are discarded. We consider a pixel part of a specularity if the intensity value is over $0.7$. 
Table~\ref{tab:spec} summarizes these results, showing that the baseline models lose a significant amount of features and inliers if we ignore specularity features (\textit{w/o S}), confirming the suspicion that they fire too much on specularities in this environment. 
%we measure how many features were actually detected in specularity regions (pixels where grayscale value is greater or equal than $180$). 
%Table~\ref{tab:spec} summarizes these results. Both SIFT and SuperPoint Base suffer similar reductions (into half), confirming the suspicion that they focus too much on specularities in this environment. 
In contrast, the proposed E-SuperPoint effectively removes reliance in specularities and allows the detector to focus on other image patterns which have higher chances of being stable. %The inlier number obtained for E-SuperPoint+S is higher than for other methods, even if the initial number of features extracted is lower. %Most methods process the whole image without discriminating the parts where there is mostly specularity 
%``noise'', while E-SuperPoint+S tends to avoid it. 
Fig.~\ref{fig:inliers1} shows several matching examples.  Higher resolution version of them and additional examples can be found in the supplementary material. 
%and video.
%, Figure 3 and video, for better visualization. 

\paragraph{Rotation estimation from matches.}
%Since we computed the pseudo-ground truth of the Eval set with COLMAP, in this experiment 
%The movement of the camera in colonoscopy sequences is mostly forwards/backwards, which results in a very narrow baseline that hinders translation estimation. But estimating camera rotation is doable. The pseudo-ground truth (pGT) pose estimation is used to compute rotation estimation errors. For every image pair 1 second apart, we estimate the Essential Matrix with RANSAC and compare the rotation it provides with respect to the pGT estimated rotation.
We compare the RANSAC-estimated essential matrix and the pseudo-ground truth essential matrix to compute the rotation estimation error for every pair of images. These values are summarized in Tab.~\ref{tab:rotation_error}, where we show the values of relevant percentiles of the errors obtained. E-SuperPoint achieves lower rotation error than all the other methods. Additionally, when counting the percentage of the estimations that obtain an error lower than $30$ degrees, E-SuperPoint succeeds $\textbf{71.6}\%$ of the time, while the second best, SuperPoint Base, only $61.3\%$. 
%Our models behave the best, with similar median metrics and less failure cases (i.e, cases with error $>30^{\circ}$). 
%Images with many reflections (specularities) correspond to one type of failure case. Unfortunately, it is not frequent in our benchmark \textit{test set} by construction: this set is composed by frames successfully used by COLMAP on the reference 3D reconstruction using SIFT features; but SIFT fails in frames with many reflections. 
Figure~\ref{fig:inliers1} shows some examples where we can see that E-SuperPoint is more robust than other methods: 
Top row example shows how E-SuperPoint ignores the specularities in the image completely (there are no features extracted on top of it), and the bottom row example shows that our model finds more matches and better spread over the image than the other methods.
%even E-Superpoint fails while E-SuperPoint+S error is $6.02^{\circ}$. %$15.05^{\circ}$ and , respectively.
%Besides, we have observed that E-SuperPoint+S can deal better with images full of reflections, such as the examples in  
%Figure~\ref{fig:inliers1}, where  E-Superpoint fails while  E-SuperPoint+S errors are $15.05^{\circ}$ and $6.02^{\circ}$, respectively. These images full of reflections are underrepresented in the sub-sequences successfully reconstructed by COLMAP used as pseudo-ground truth, and therefore in our evaluation set. This is because SIFT fails in those cases.
%see the last interval in  Figure~\ref{fig:rotation_error}(b) plot.
%as shown in the supplementary material, section \ref{sec:matching_examples}. 
%The rotation error (in degrees) for Figure~\ref{fig:inliers1} examples is of $140.557$ and $164.513$ for E-Superpoint, while 
%This confirms the benefits of avoiding specularity points to achieve more accurate pose estimations.
The improvement is largely due to the adaptation we made to better deal with specularities for feature extraction.
\begin{table}[!tb]
\centering
\footnotesize
    \begin{tabular}{|l|cc|cc|cc|}
    \cline{2-5}
     \multicolumn{1}{c}{} & \multicolumn{2}{|c|}{\textbf{Feat/Img}} & \multicolumn{2}{|c|}{\textbf{E Inliers}} \\
     \multicolumn{1}{c|}{} & all feat. & w/o S & all feat. & w/o S \\
    \hline
    % $2350.2$ & $0.0/230.1$ ($0.0\%$) & $nan$ & $nan$ & $148.2/230.1$ ($64.4\%$) & $0.268$ & $11.9$ & $124.3/230.1$ ($54.0\%$) & $0.268$ & $7.7$ & $71.6/230.1$ ($31.1\%$) & $0.253$ & $9.3$
    % $2006.2$ & $0.0/199.2$ ($0.0\%$) & $nan$ & $nan$ & $129.1/199.2$ ($64.8\%$) & $0.266$ & $11.7$ & $106.6/199.2$ ($53.5\%$) & $0.265$ & $7.5$ & $62.9/199.2$ ($31.6\%$) & $0.249$ & $9.1$
    \textbf{SIFT} & $2350.2$ & $2006.2$ ($85.4\%$) & $148.2$ & $129.1$ ($87.1\%$)\\
    % $2163.0$ & $0.0/1005.2$ ($0.0\%$) & $nan$ & $nan$ & $153.0/1005.2$ ($15.2\%$) & $0.260$ & $8.5$ & $118.6/1005.2$ ($11.8\%$) & $0.260$ & $5.8$ & $80.9/1005.2$ ($8.0\%$) & $0.247$ & $6.0$
    % $867.7$ & $0.0/405.0$ ($0.0\%$) & $nan$ & $nan$ & $61.8/405.0$ ($15.3\%$) & $0.252$ & $6.1$ & $41.3/405.0$ ($10.2\%$) & $0.249$ & $3.6$ & $35.2/405.0$ ($8.7\%$) & $0.244$ & $4.3$
    \textbf{ORB} & $2163.0$ & $867.7$ ($40.1\%$) & $153.0$ & $61.8$ ($40.4\%$)\\
    \hline
    % $1333.7$ & $0.0/577.8$ ($0.0\%$) & $nan$ & $nan$ & $96.4/577.8$ ($16.7\%$) & $0.268$ & $11.4$ & $79.1/577.8$ ($13.7\%$) & $0.265$ & $8.3$ & $57.1/577.8$ ($9.9\%$) & $0.259$ & $8.2$
    % $1035.2$ & $0.0/453.6$ ($0.0\%$) & $nan$ & $nan$ & $73.5/453.6$ ($16.2\%$) & $0.263$ & $10.6$ & $59.1/453.6$ ($13.0\%$) & $0.260$ & $7.4$ & $43.9/453.6$ ($9.7\%$) & $0.253$ & $7.6$
    \textbf{SP Base} & $1333.7$ & $1035.2$ ($77.6\%$)  & $96.4$ & $73.5$ ($76.2\%$)\\
    % $4500.9$ & $0.0/1349.7$ ($0.0\%$) & $nan$ & $nan$ & $278.9/1349.7$ ($20.7\%$) & $0.295$ & $13.2$ & $197.5/1349.7$ ($14.6\%$) & $0.272$ & $11.0$ & $172.0/1349.7$ ($12.7\%$) & $0.294$ & $9.8$
    % $4431.3$ & $0.0/1329.8$ ($0.0\%$) & $nan$ & $nan$ & $274.7/1329.8$ ($20.7\%$) & $0.294$ & $13.1$ & $195.4/1329.8$ ($14.7\%$) & $0.272$ & $10.9$ & $169.1/1329.8$ ($12.7\%$) & $0.293$ & $9.7$
    \textbf{E-SP} & $4500.9$ & $4431.3$ ($\textbf{98.5}\%$) & $278.9$ & $274.7$ ($\textbf{98.5}\%$)\\
    \hline
    \end{tabular}
    \caption{Influence of specularities in the matching results (in EndoMapper \textit{test set}). all feat: total number; w/o S: number without features that fall into specularities.}
    \label{tab:spec}
\end{table}

% Errors figure 2
%row 2 col 4 E-SP 140.557
%row 2 col 5 E-SP+S 15.051
%row 4 col 4 E-SP 164.513
%row 4 col 5 E-SP+S 6.018

\begin{table}[!tb]
\centering
\footnotesize
    \begin{tabular}{|l|ccccc|}
    \cline{2-6}
     %\multicolumn{1}{c}{} & %\multicolumn{5}{|c|}{\textbf{Rotation estimation error}} \\
    \cline{2-6}
     \multicolumn{1}{c|}{Percentile} & 10$^{th}$ & 20$^{th}$ & 30$^{th}$ & 40$^{th}$ & median\\
    \hline
    % 6sift_o3_c10_e10_s16_r0.8_0_1_40_5.npy 3.1741216805659955 5.500965366188679 8.423371881605748 12.37668002727012 20.134250525549234 0.5709322524609148
    \textbf{SIFT} & $3.2$ & $5.5$ & $8.4$ & $12.4$ & $20.1$\\
    % 6orb_sf12_nl8_5_1_40_5.npy 3.7349368488000927 7.265262709932823 13.427173654513004 27.433336498726682 45.76946655122299 0.4147365373480023
    \textbf{ORB} & $3.7$ & $7.3$ & $13.4$ & $27.4$ & $45.8$\\ 
    \hline
    % 6super_0_1_40_5.npy 3.198049064115121 5.789868340202082 9.197361747828658 13.373503181265324 19.760419958442768 0.6130573248407644
    \textbf{SP Base} & $3.2$ & $5.8$ & $9.2$ & $13.4$ & $19.8$\\ 
    % 6specga_9-4_d3_0_1_40_5.npy 2.6490167441433083 4.84603235138351 7.465624575906415 10.461832712893106 14.451734665933088 0.7164157498552403
    \textbf{E-SP} & $\textbf{2.6}$ & $\textbf{4.8}$ & $\textbf{7.5}$ & $\textbf{10.5}$ & $\textbf{14.5}$\\
    \hline
    \end{tabular}
    \caption{Rotation estimation error for pairs of frames 1 second apart. Percentile values of the errors obtained by each method.}
    \label{tab:rotation_error}
\end{table}

\section{Conclusions}
This work\footnote{This project has been funded by the European Union’s Horizon 2020 research and innovation programme under grant agreement No 863146 and Aragon Government FSE-T45\_20R.} studies the performance of local features in \textit{in the wild} endoscopic environments, using data captured during routine medical practice. We compare the effectiveness of hand-crafted local features against deep learning-based ones, in particular SIFT, ORB and SuperPoint. Although hand-crafted features are still a dominant choice in this field, we show how deep learning based features can surpass them in the considered challenging environments. Besides, we have trained and adapted the general-purpose SuperPoint to better fit the challenges of endoscopic imagery.
Our evaluation, on endoscopies of different patients, is focused on the quality of the recovered 3D camera motion. % showing that promising results in the early stages does not always imply a better performance later on. Since learned features have a tendency of working at lower image resolutions, we have proposed segmenting the image and we have shown that doing this brings the method up to par with the other multi-scale methods. 
Our results show that SuperPoint adaptation provides  more numerous and non-specular features, and more disperse correspondences, essential for accurate and robust 3D geometry estimations. 
%Besides, a qualitative comparison shows how the final 3D reconstruction using SuperPoint appears to be similar or better than those obtained with SIFT in the well-known COLMAP 3D reconstruction software.
% evaluated state-of-the-art in feature learning using endoscopic ...
% proposed a modification that behaves better and release the model

% Acknowledgments---Will not appear in anonymized version
%\midlacknowledgments{We thank a bunch of people.}

%Bibliography
\bibliographystyle{splncs04}
\bibliography{arxiv-fullpaper.bib}  
\newpage
\appendix

\section{Supplementary material}
\begin{figure}[!h]
    \centering
    \footnotesize 
    \begin{tabular}{@{}c@{\hspace{0.9mm}}c@{\hspace{0.9mm}}c@{\hspace{0.9mm}}c@{\hspace{0.9mm}}c@{\hspace{0.9mm}}c|c}
        % \multicolumn{3}{c}{\footnotesize Hamlyn} & 
         \multicolumn{6}{c|}{\footnotesize Endoscopic Video examples} & \textit{Mask}\\
        %HAMLYN
        % \includegraphics[width=0.10\textwidth]{images/calibrated05_L_0063.png} & 
        % \includegraphics[width=0.10\textwidth]{images/calibrated16_L_0089.png} & 
        % \includegraphics[width=0.10\textwidth]{images/calibrated17_L_0023.png} &
        %HCULB
        \includegraphics[width=0.12\textwidth]{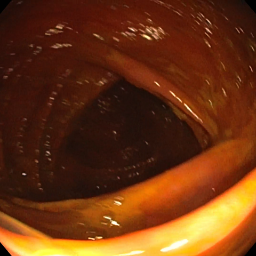} &
        \includegraphics[width=0.12\textwidth]{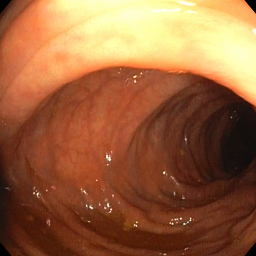} &
        \includegraphics[width=0.12\textwidth]{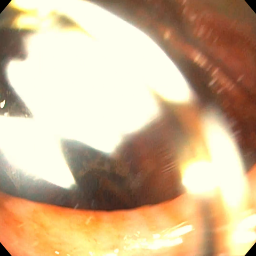} &
        %\includegraphics[width=0.2\textwidth]{images/calibrated16_L_0089.png} & & \includegraphics[width=0.2\textwidth]{images/HCULB_00039_procedure_lossy_h264_0072.png}\\
        %\includegraphics[width=0.2\textwidth]{images/calibrated17_L_0023.png} & & \includegraphics[width=0.2\textwidth]{images/HCULB_00051_procedure_lossy_h264_0091.png}\\
        %WEISS
        \includegraphics[width=0.12\textwidth]{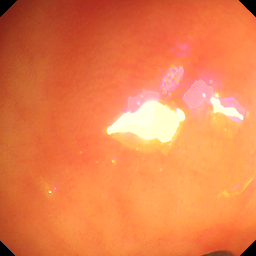} &
        \includegraphics[width=0.12\textwidth]{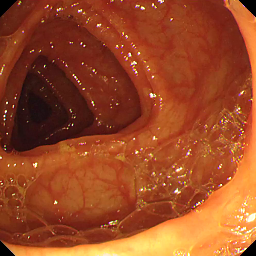} &
        \includegraphics[width=0.12\textwidth]{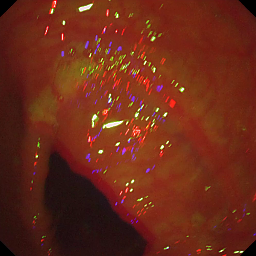} & \includegraphics[width=0.12\linewidth]{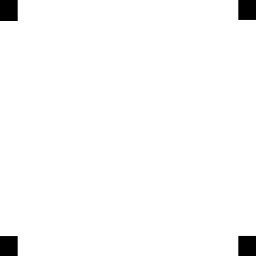}\\
    \end{tabular}
    \caption{\textbf{Sample frames from \textit{Train set}.} They often present poor quality frames and many artifacts. We apply a simple mask (\textit{Mask}) to ignore black corners, which are a constant artifact due to the endoscope camera type.}%  More sample images in the supplementary video.
    \label{fig:dataset_examples}
\end{figure}

\begin{figure}[!h]
    \centering
    \footnotesize 
    \begin{tabular}{@{}c@{\hspace{0.9mm}}c@{\hspace{0.9mm}}c|c@{\hspace{0.9mm}}c@{\hspace{0.9mm}}c|c}
        % \multicolumn{3}{c}{\footnotesize Hamlyn} & 
         \multicolumn{3}{c|}{\footnotesize Hyper-Kvasir frames} & \multicolumn{3}{c|}{\footnotesize Cropped frames} & \textit{Mask}\\
        \includegraphics[width=0.13\textwidth]{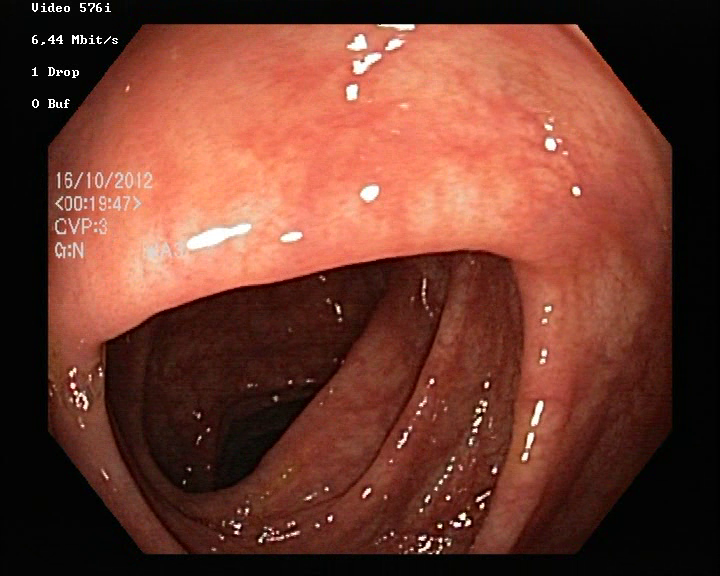} &
        \includegraphics[width=0.13\textwidth]{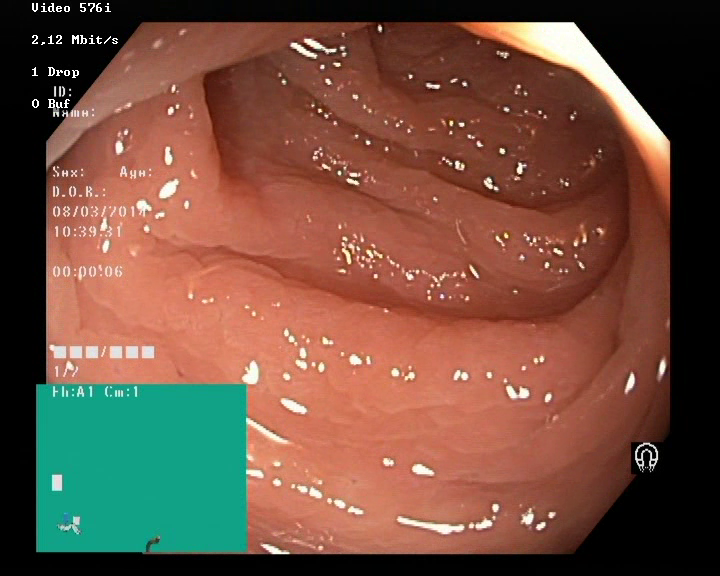} &
        \includegraphics[width=0.13\textwidth]{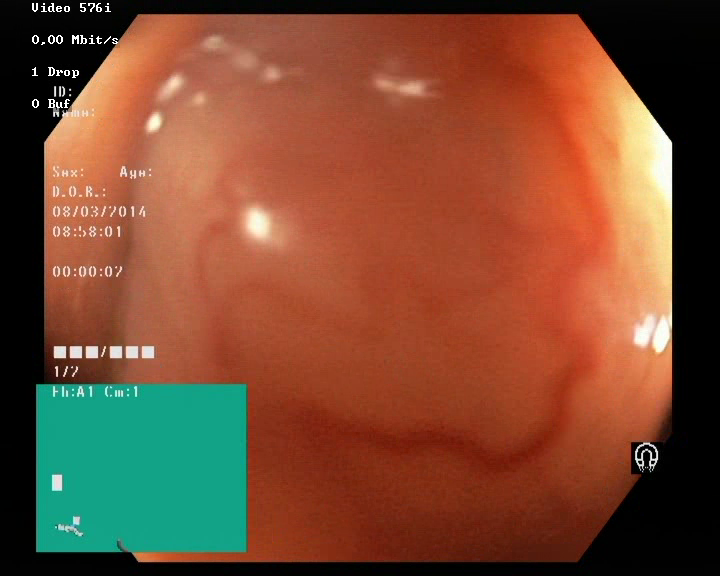} &
        
        \includegraphics[width=0.12\textwidth]{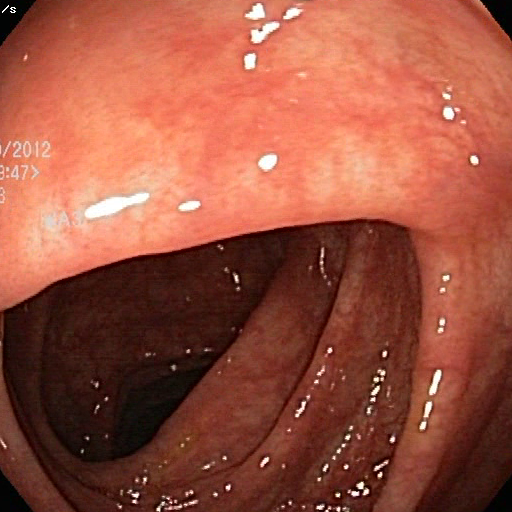} &
        \includegraphics[width=0.12\textwidth]{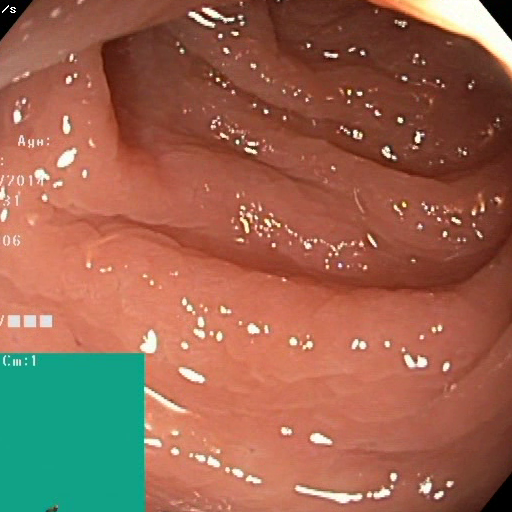} &
        \includegraphics[width=0.12\textwidth]{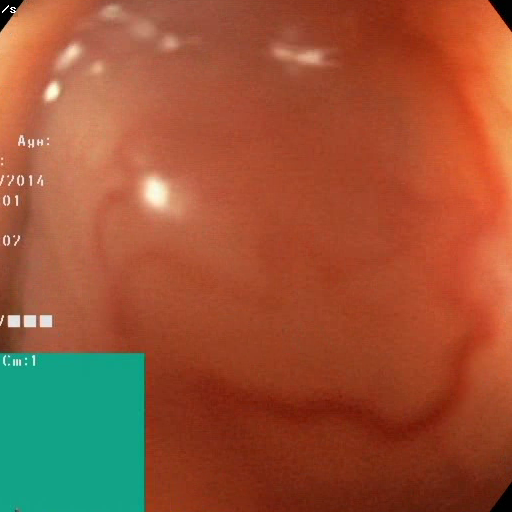} &
        \includegraphics[width=0.12\linewidth]{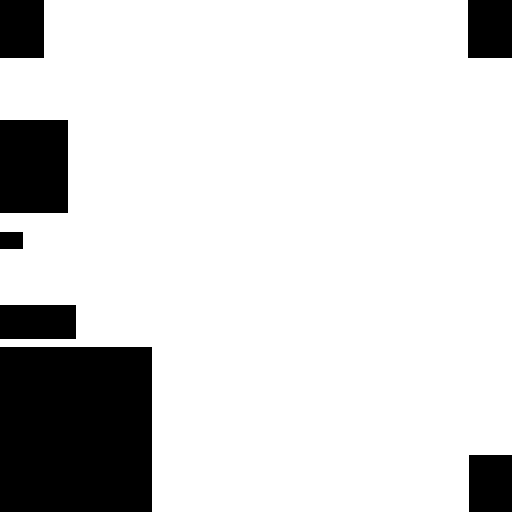}\\
    \end{tabular}
    \caption{\textbf{Sample frames from Hyper-Kvasir~\cite{borgli2020hyperkvasir}}, the labeled videos in ``lower-gi-tract/quality-of-mucosal-view/BBPS-2-3''. They are significantly different from \textit{Train set} and EndoMapper datasets, since they include additional augmented information superimposed to the left of the frames. The \textit{Mask} allows us to ignore these parts and the black corners.}% More sample images in the supplementary video.
    \label{fig:dataset_hyperkvasir}
\end{figure}

\begin{table*}[!h]
    \centering
    \small
    \begin{tabular}{|c|c|c|c|c|c|}
    \cline{2-6}
    \multicolumn{1}{c|}{} & \textbf{Impl.} & \textbf{Max. Feat} & \textbf{Matching} & \textbf{Distance} & \textbf{Other parameters}\\
    \hline
    \textbf{SIFT} & OpenCV & $10000$ & Ratio ($0.8$) & L2 Norm & Contrast threshold$=0.1$\\%~\cite{lowe2004distinctive}
    \textbf{ORB} &  OpenCV & $10000$ & BFMatcher & Hamming & -\\%~\cite{rublee2011orb}
    \textbf{SuperPoint} & \cite{jau2020deep} & $10000$ & BFMatcher & L2 Norm & -\\%~\cite{detone2018superpoint}
    \hline
    \end{tabular}
    \caption{\textbf{Implementation details.} For all our experiments we use OpenCV 3.4.2. RANSAC algorithm is set with confidence threshold of $0.9999$, and all matches within a distance of $3$ pixels of their matched pair when reprojecting are considered inliers.}
    \label{tab:metrics}
\end{table*}

\begin{figure*}[!tb]
    \centering
    \includegraphics[width=0.85\textwidth]{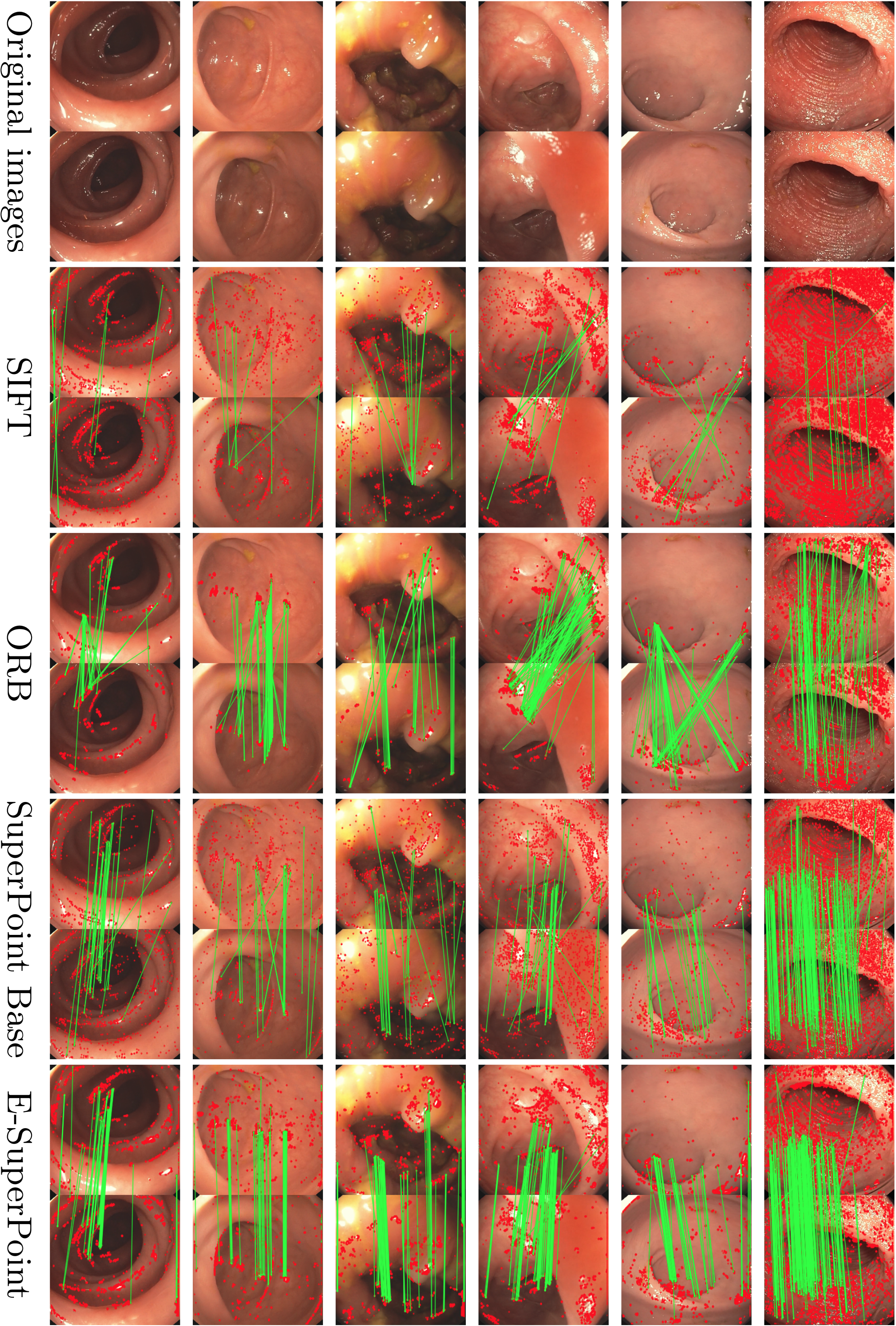}% trim={0 7cm 0 0},clip,
    \caption{\textbf{Additional examples in higher resolution} of features (red circles) and matches (green lines) obtained in pairs of 1 second apart frames from \textbf{EndoMapper~\cite{azagra2022endomapper}} dataset. As discussed in the main paper, our E-Superpoint achieves more reliable correspondences, more spread than the rest, and is able to mitigate the amount of features located in specularities.}%  More examples in the supplementary video.
    \label{fig:inliers1_big}
\end{figure*}

\end{document}